\documentclass[twocolumn]{aastex631}
                           
\shorttitle{}
\shortauthors{Ma et al.}
\graphicspath{{./}{figures/}}

\usepackage{appendix}
\usepackage{amsmath}
\usepackage{enumitem}
\usepackage{graphicx}
\usepackage{wrapfig}
\usepackage{amssymb}
\usepackage{threeparttable}
\usepackage[ruled,vlined]{algorithm2e}
\begin{document}

\title{Towards optimal photometric calibration of digital astronomical plates with deep learning}
\correspondingauthor{Haibo Yuan}
\email{yuanhb@bnu.edu.cn}

\author[0009-0003-1069-1482]{Mingyang Ma}
\affiliation{Institute for Frontiers in Astronomy and Astrophysics, Beijing Normal University, Beijing, 102206, China}
\affiliation{School of Physics and Astronomy, Beijing Normal University, Beijing, 100875, China} 

\author[0000-0003-2471-2363]{Haibo Yuan}
\affiliation{Institute for Frontiers in Astronomy and Astrophysics, Beijing Normal University, Beijing, 102206, China}
\affiliation{School of Physics and Astronomy, Beijing Normal University, Beijing, 100875, China} 

\author[0000-0002-9824-0461]{Lin Yang}
\affiliation{Department of Cyber Security, Beijing Electronic Science and Technology Institute, Beijing, 100070, China}

\author[0000-0001-8424-1079]{Kai Xiao}
\affiliation{School of Astronomy and Space Science, University of Chinese Academy of Sciences, Beijing 100049, People's Republic of China}

\author[0000-0002-1259-0517]{Bowen Huang}
\affiliation{Institute for Frontiers in Astronomy and Astrophysics, Beijing Normal University, Beijing, 102206, China}
\affiliation{School of Physics and Astronomy, Beijing Normal University, Beijing, 100875, China} 

\author{Shiyin Shen}{}
\affiliation{Shanghai Astronomical Observatory, Chinese Academy of Sciences, Shanghai 200030, China}
\author{Zhengjun Shang}{}
\affiliation{Shanghai Astronomical Observatory, Chinese Academy of Sciences, Shanghai 200030, China}
\author{Yong Yu}{}
\affiliation{Shanghai Astronomical Observatory, Chinese Academy of Sciences, Shanghai 200030, China}
\affiliation{School of Astronomy and Space Science, University of Chinese Academy of Sciences, Beijing 100049, China}
\author{Meiting Yang}{}
\affiliation{Shanghai Astronomical Observatory, Chinese Academy of Sciences, Shanghai 200030, China}
\author{Zhenghong Tang}{}
\affiliation{Shanghai Astronomical Observatory, Chinese Academy of Sciences, Shanghai 200030, China}
\affiliation{School of Astronomy and Space Science, University of Chinese Academy of Sciences, Beijing 100049, China}
\author{Jianhai Zhao}{}
\affiliation{Shanghai Astronomical Observatory, Chinese Academy of Sciences, Shanghai 200030, China}

\begin{abstract}
Photometric calibration of digitized photographic plates is commonly modeled with separable magnitude-, color-, and position-dependent terms, but this separability can break down when image quality varies across the field in a magnitude-dependent way, leaving coupled spatial systematics in the residuals. We introduce a deep-learning calibration framework, the Multi-Feature Fused Network (MFF-Net), which takes instrumental magnitude, color, and pixel coordinates as input and learns a single nonlinear correction that jointly captures their coupled dependencies. Tests on 1{,}200 digitized Chinese plates show that MFF-Net consistently outperforms the MYX25 method \citep{2025ApJS..280...18M}, improving the 5th--95th percentile precision from 0.11--0.26~mag to 0.08--0.18~mag and delivering an approximately factor-of-two gain for bright sources. The learned correction largely removes the magnitude--position coupling seen in post-calibration residual maps, enabling higher-precision plate photometry and more reliable use of large historical plate archives.

\end{abstract}

\keywords{Stellar photometry, Astronomy data analysis, Calibration, Neural networks}

\section{Introduction} \label{sec:intro}
Photographic plates entered astronomical use in the 1840s, shifting the field from hand-drawn sketches to systematic, instrument-based imaging. By the 1880s they underpinned major sky surveys and, for roughly a century, served as the primary medium for optical observations until CCDs became widespread in the 1980s. More than 10 million historical plates are now preserved worldwide (\citealt{2018AN....339..408H}; \citealt{2019AN....340..690H}). Since the late 20th century, large digitization efforts---including the SuperCOSMOS Sky Survey (SSS; \cite{2001MNRAS.326.1279H}), Digital Access to a Sky Century @ Harvard (DASCH; \cite{2009ASPC..410..101G}), and the Archives of Photographic Plates for Astronomical USE (APPLAUSE; \cite{2024A&A...687A.165E})---have aimed to integrate these legacy observations into modern astronomical infrastructures. Plates can be grouped into direct-imaging and spectroscopic material; current digitization programs primarily target direct-imaging plates, and the calibration method developed here is restricted to that class.

Unlike CCDs, whose photometric response is approximately linear, photographic plates are strongly nonlinear. The characteristic curve---relating log exposure to optical density---is S-shaped: it is shallow when underexposed, approximately linear at intermediate exposures, and saturates when overexposed \citep{hurter1890}. In addition, nonuniform developer agitation and wide-field optical aberrations imprint complex, spatially varying, flat-field-like patterns on plate images.

More than a century ago, accurate all-sky catalogs and dense photometric standards were unavailable. With only a handful of standards, astronomers could estimate plate response curves only crudely and therefore relied on empirical transformations between instrumental and standard magnitudes. By the early 21st century, the UK SSS project used the Tycho Catalog \citep{1997A&A...323..620K} and GSPC1 \citep{1988ApJS...68....1L} to construct an all-sky standard-star sample down to $V\sim15$ mag for plate calibration \citep{2001MNRAS.326.1295H}. To push the calibration to the plate limits, they imposed a linear constraint beyond the standard-star range using $B\sim21$ mag CCD data from 40 high-latitude fields \citep{1995MNRAS.276...33B,1999MNRAS.306..592C}. When the DASCH project began, the deeper all-sky APASS survey \citep{2016yCat.2336....0H} (to $V\sim17$ mag) provided sufficient density and depth to cover its plates. This enabled explicit modeling of detector response curves to reduce brightness-dependent systematics and use of the large standard-star sample to map and correct complex flat-field structure \citep{2010AJ....140.1062L}. The German APPLAUSE project adopted a similar strategy \citep{2024A&A...687A.165E}, using Gaia EDR3 \citep{2021A&A...650C...3G} as standards. Gaia DR3 \citep{2023A&A...674A...1G} further added over 200 million low-resolution BP/RP spectra (XP; $330\,\mathrm{nm}\le\lambda\le1050\,\mathrm{nm}$), mostly with $G<17.65$. After correcting magnitude- and color-dependent systematics \citep{2024ApJS..271...13H}, \cite{2026NatSD..13..265X} performed synthetic photometry in more than 200 bands with $<10\,\mathrm{mmag}$ precision and incorporated these magnitudes into the BEst Star (BEST) database (\url{https://nadc.china-vo.org/data/best/}). Using $UBVRI$ bands from BEST, \cite{2025ApJS..280...18M} built standard-star samples to calibrate China’s digitized plates \citep{2024RAA....24e5010S} by modeling the plate photometric zero point as separable functions of magnitude, color, and field position (Equation~\ref{eq:1}):
\begin{align}
m_{\mathrm{corr}} = m_{\mathrm{inst}} + f_m(m_{\mathrm{inst}}) + f_c(BP-RP) + f_p(x,y)
\label{eq:1}
\end{align}
Here, $m_{\mathrm{inst}}$ and $m_{\mathrm{corr}}$ are the magnitudes before and after calibration, respectively; $BP$ and $RP$ are from Gaia DR3; and $x$ and $y$ are stellar pixel coordinates. The terms $f_m$, $f_c$, and $f_p$ represent the magnitude, color, and flat-field components. Using this model, \citeauthor{2025ApJS..280...18M} (\citeyear{2025ApJS..280...18M}; hereafter MYX25) calibrated the catalog to the JKC photometric system \citep{2012PASP..124..140B}, achieving a typical precision of 0.1--0.2 mag.

Overall, photometric calibration of photographic plates has advanced substantially. Nevertheless, many widely used approaches treat the dominant systematics as separable terms \citep{2001MNRAS.326.1295H, 2010AJ....140.1062L, 2024A&A...687A.165E, 2025ApJS..280...18M}, even though these effects can be coupled in practice. In our recent analysis of the MYX25 residuals, we find strong coupling, particularly between source brightness and spatial position.

\cite{2013PASP..125..857T} introduced a local calibration scheme that corrects each target using nearby reference stars of similar magnitude, and demonstrated it on DASCH plates. While effective, the approach requires a sufficient number of suitable neighbors: in sparse regimes (especially for bright stars), the magnitude and/or spatial search window must be enlarged to meet minimum-count requirements, reducing locality. In addition, discretizing the (magnitude, position) space into non-overlapping bins can produce artificial discontinuities at bin boundaries.

To address these limitations, we model the dominant factors simultaneously and continuously. Specifically, we introduce a deep-learning framework for photometric calibration of digitized plates based on a Multi-Feature Fused Network (MFF-Net; Equation~\ref{eq:2}):
\begin{align}
m_{\mathrm{corr}} = m_{\mathrm{inst}} + f_{\mathrm{MFF}}\bigl(m_{\mathrm{inst}},\,BP-RP,\,x,\,y\bigr)
\label{eq:2}
\end{align}
With strong nonlinear approximation capability and efficient handling of high-dimensional inputs, MFF-Net is well suited to jointly modeling and correcting coupled systematics. Applied to digitized Chinese astronomical plates, it reduces coupling between photometric and spatial variables and achieves higher precision than the MYX25 method.

The paper is organized as follows. Section~\ref{sec:MYX25} reviews the limitations of the MYX25 approach. Section~\ref{sec:network} presents our deep learning-based calibration method, including the network architecture, training strategy, and illustrative examples. Section~\ref{sec:conclusion} summarizes the main results and outlines future work.

\section{Limitations of the MYX25 approach} \label{sec:MYX25}

Analysis of calibrated digitized plates from China, particularly wide-field plates, highlights a key limitation of the MYX25 method: post-calibration residuals retain systematic, magnitude-dependent spatial structure, indicating that photometric response and field position are not fully decoupled.

\begin{figure*}[ht!] \centering
\includegraphics[width=15cm]{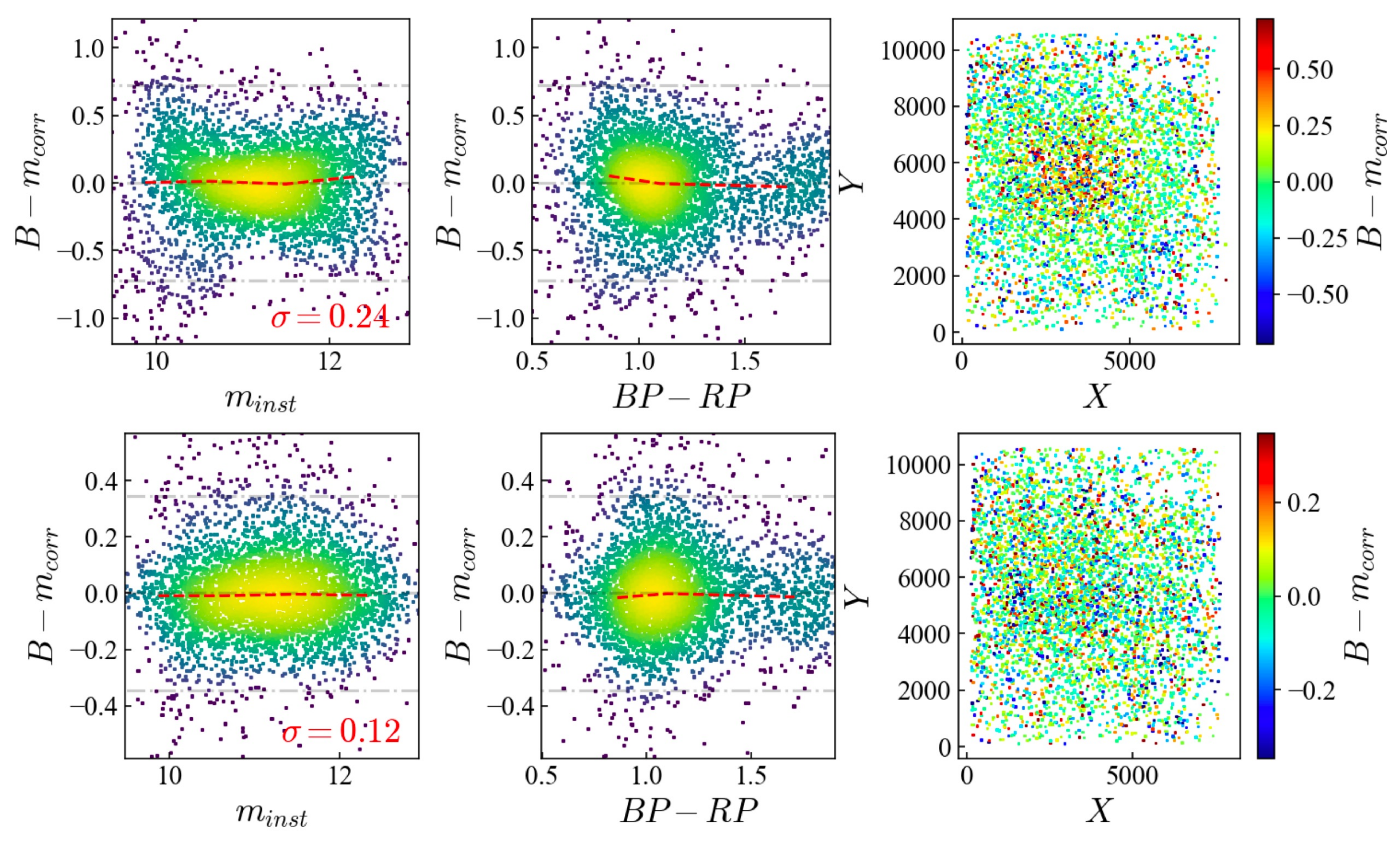}
\caption{Calibration residuals for plate ZT5810N973001 as functions of instrumental magnitude, color, and pixel coordinates.
The upper row shows results from MYX25, with a larger overall error of 0.24~mag; the lower row shows results from MFF-Net, reducing the error to 0.12~mag.}
\label{Fig:residual}
\end{figure*}

\begin{figure*}[ht!] \centering
\includegraphics[width=18cm]{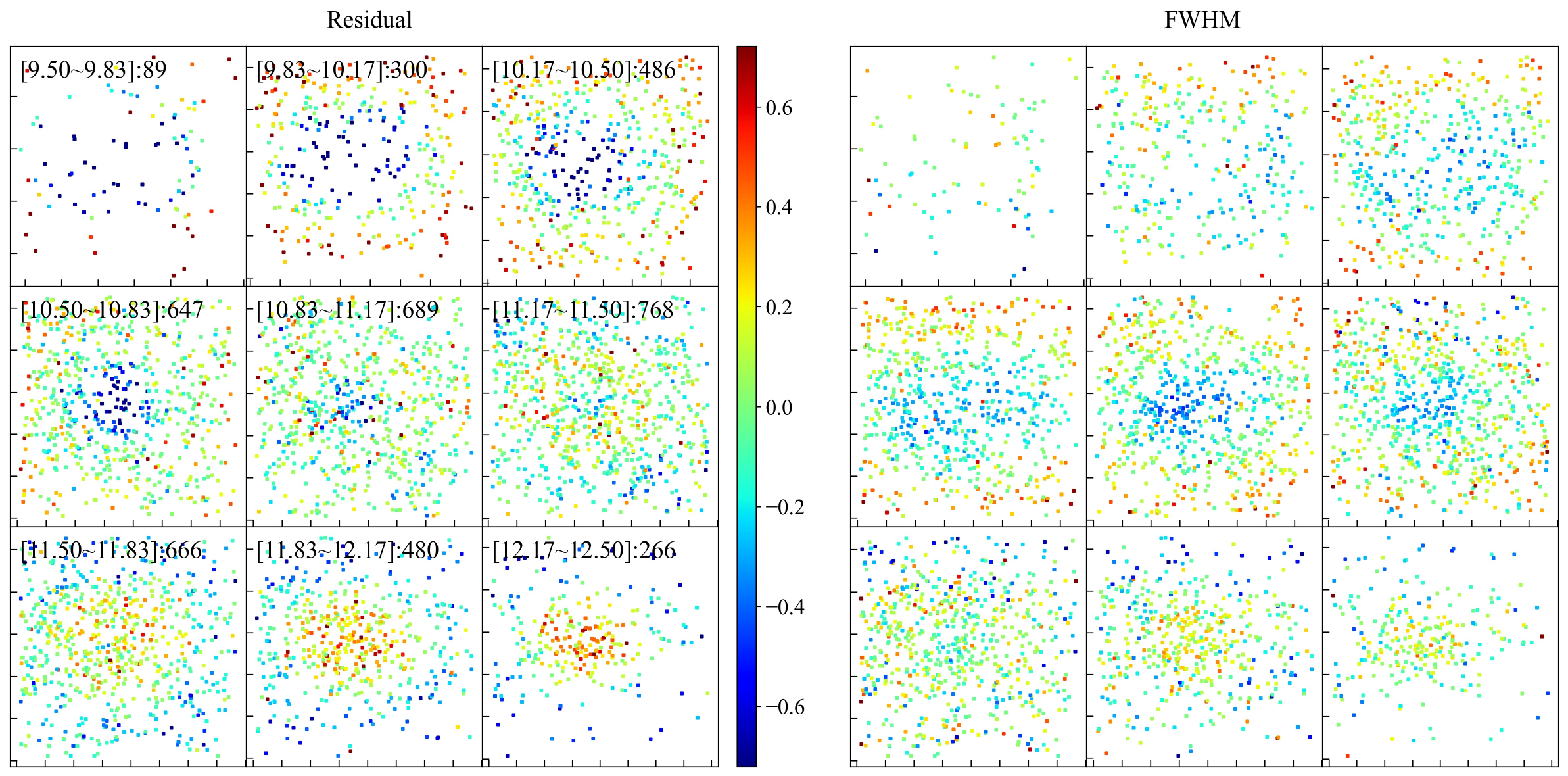}
\caption{MYX25 calibration residuals (left) and measured FWHM (right) for plate ZT5810N973001, shown in nine magnitude bins. The magnitude range and number of stars are labeled in the left panels, which share a common color scale. The right panels use individual color scales to highlight spatial structure (red: higher values; blue: lower values).}
\label{Fig:couple_mrmt}
\end{figure*}

Fig.~\ref{Fig:residual} (upper row) shows residuals versus instrumental magnitude, color, and position for plate ZT5810N973001 (from Purple Mountain Observatory; field of view $3.4\,\mathrm{deg}^2$). While the residuals appear broadly uniform when viewed in aggregate, splitting the sample into nine magnitude bins reveals a clear annular pattern within each bin (left panel of Fig.~\ref{Fig:couple_mrmt}).

We argue that this effect is driven by spatial variations in image quality, quantified by the full width at half maximum (FWHM; Fig.~\ref{Fig:couple_mrmt}, right panel). Because the plate response depends on surface brightness rather than total flux, two stars with the same total brightness but different FWHM have different surface brightness and therefore require different magnitude-term corrections on the nonlinear characteristic curve. MYX25, however, applies a uniform correction based only on total brightness, introducing a systematic error coupled to FWHM; because FWHM is often position-dependent, this produces the annular patterns seen in the residual maps.

To test this hypothesis, we analyzed additional plates. When the image quality is spatially uniform, the residual maps remain correspondingly uniform across magnitude. By contrast, plates with spatially varying image quality often show magnitude--position coupling in the residuals, although the strength of the effect varies from plate to plate. Moreover, we find that magnitude‑dependent spatial structures in the residuals occur exclusively in plates for which the spatial pattern of image quality itself varies with magnitude. We hypothesize that the impact of plate image quality on the photometric system, to the extent that it is independent of magnitude, can be incorporated into the flat‑field term. On this basis, we developed and conducted two simulation experiments to evaluate this hypothesis.

\begin{figure*}[ht!] \centering
\includegraphics[width=18cm]{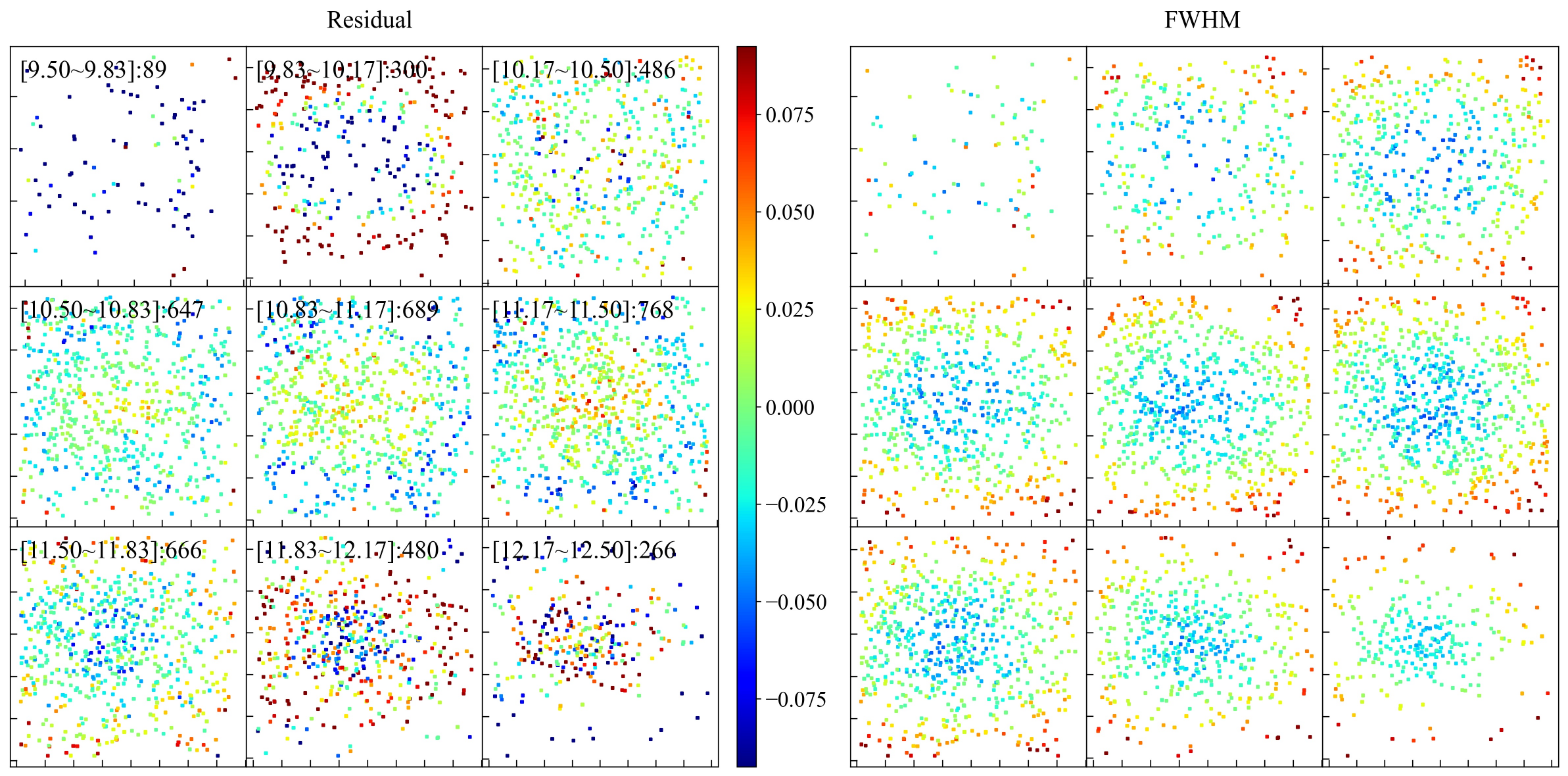} 
\caption{
MYX25 calibration residuals (left) and FWHM of the simulated data from Experiment 1, plotted as in Fig.~\ref{Fig:couple_mrmt}.}
\label{Fig:couple_smlt_fwhm}
\end{figure*}

Both experiments used plate ZT5810N973001. In each case, we fixed the standard magnitudes (true total flux) and pixel coordinates of the BEST standard stars. In Experiment~1, we assigned FWHM values using an empirical prescription in which FWHM increases toward brighter magnitudes and with increasing distance from the field center (Fig.~\ref{Fig:couple_smlt_fwhm}, right panel). In Experiment~2, we adopted the plate’s measured FWHM values (Fig.~\ref{Fig:couple_mrmt}, right panel), for which image quality is better in the central region for stars brighter than 11.5~mag but better in the outer region for stars fainter than 11.5~mag. Because photographic plates respond to surface brightness rather than total flux, we converted each star’s standard magnitude to a surface-brightness-normalized flux to account for FWHM-related biases. We then used the plate response curve to transform these values into simulated instrumental magnitudes.

\begin{figure}[ht!] \centering
\includegraphics[width=8.5cm]{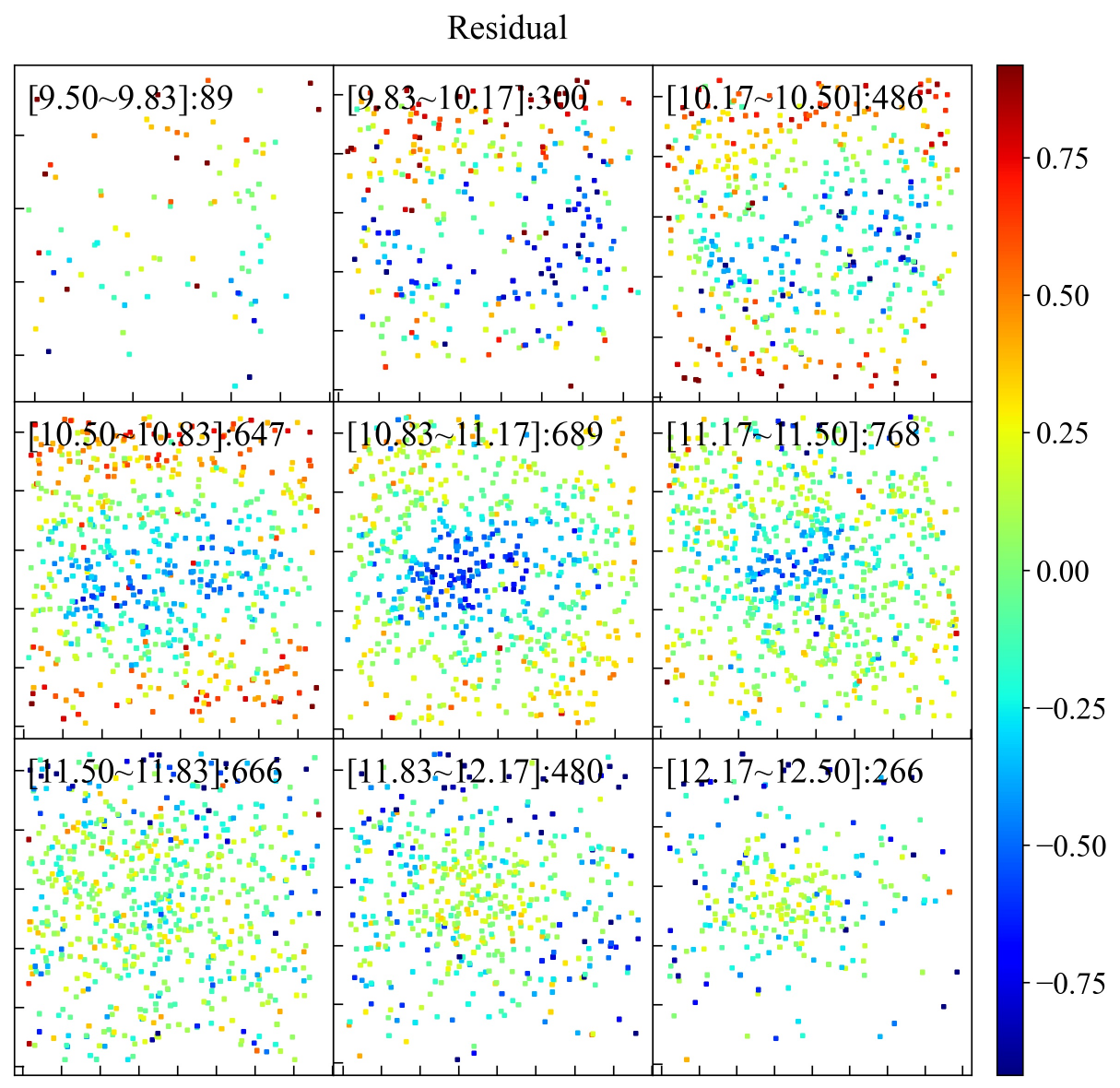} 
\caption{MYX25 calibration residuals of the simulated data from Experiment 2, plotted as in the left panel of Fig.~\ref{Fig:couple_mrmt}.}
\label{Fig:couple_smlt_minst}
\end{figure}

We then calibrated the simulated data sets with the MYX25 method. Comparing the two experiments isolates the origin of the annular residuals: when the simulated FWHM pattern is independent of magnitude, the flat-field term absorbs the FWHM-induced systematic error and the photometric precision improves markedly (Fig.~\ref{Fig:couple_smlt_fwhm}, left). By contrast, when we preserve the magnitude-dependent FWHM structure of the real plate, both the photometric precision and the residual pattern closely reproduce those of the observations (Fig.~\ref{Fig:couple_smlt_minst}).

These results indicate that the coupled magnitude--position residuals are driven by magnitude-dependent spatial variations in image quality. Such behavior is difficult to capture with models that assume separable, independent terms, motivating a calibration approach that can jointly learn nonlinear couplings among multiple parameters.

\section{A deep learning based method} \label{sec:network}

To capture couplings among brightness, color, and plate position, we develop a deep learning--based photometric calibration method based on MFF-Net, which exploits the flexibility of neural networks to learn and correct these coupled systematics.

\begin{figure}[ht!] \centering
\includegraphics[width=8cm]{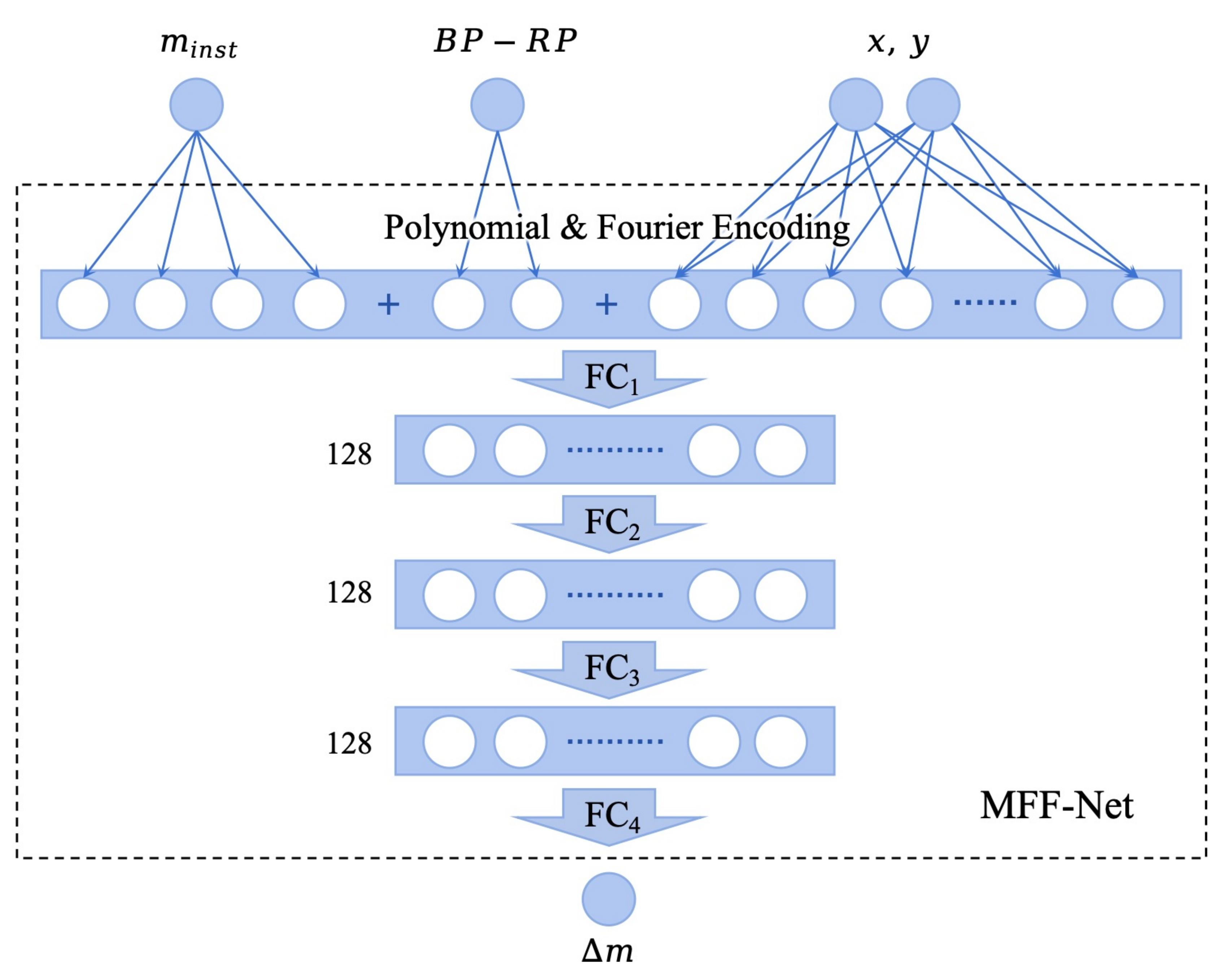} 
\caption{MFF-Net architecture.}
\label{Fig:MFF-Net}
\end{figure}

Fig.~\ref{Fig:MFF-Net} shows the MFF-Net architecture. The network ingests four features---instrumental magnitude ($m_{\mathrm{inst}}$), color ($BP-RP$), and pixel coordinates ($x, y$)---each linearly scaled to $[0,1]$. It predicts the magnitude correction $\Delta m = m_{\mathrm{std}} - m_{\mathrm{inst}}$, and the calibrated magnitude is computed as $m_{\mathrm{inst}}+\Delta m$.

The first internal layer performs feature encoding. For instrumental magnitude and color, we apply a polynomial expansion---fourth order in magnitude and second order in color---matching the MYX25 formulation. Specifically,
\begin{align}
\mathbf{p} = \bigl[ &m_{\mathrm{inst}},\; m_{\mathrm{inst}}^2,\; m_{\mathrm{inst}}^3,\; m_{\mathrm{inst}}^4,\;\nonumber\\ &(BP-RP),\; (BP-RP)^2\bigr]^{\mathsf{T}}.
\label{eq:3}
\end{align}

For pixel coordinates, we use Fourier feature encoding \citep{2020arXiv200308934M} to represent spatial systematics over multiple scales:
\begin{align}
\boldsymbol{\gamma}(x, y) = \bigl[x,\ y\bigr] \;\oplus\; \bigoplus_{k=0}^{N} \bigl[ &\sin(\pi 2^k x),\ \cos(\pi 2^k x), \nonumber\\
&\sin(\pi 2^k y),\ \cos(\pi 2^k y) \bigr],
\label{eq:4}
\end{align}
where $\oplus$ denotes vector concatenation. This smooth, differentiable encoding is well suited to gradient-based optimization and allows the network to model spatial systematics over multiple scales, capturing complex flat-field structure.
Here, $N$ is the number of frequencies: increasing $N$ enables the network to resolve finer spatial structure within the normalized coordinate domain $[0,1]$. We set $N$ based on $S$, the number of standard stars on a plate (Eq.~\ref{eq:5}):
\begin{align}
N =
\begin{cases}
2, & 1 \times 10^3 \leq S < 5 \times 10^3,\\
3, & 5 \times 10^3 \leq S < 5 \times 10^4,\\
4, & S \geq 5 \times 10^4.
\end{cases}
\label{eq:5}
\end{align}

Following \cite{2021iccv.conf..580B}, we retain the raw coordinates $(x,y)$ alongside the Fourier encoding, which helps capture large-scale (approximately linear) spatial trends and improves extrapolation.
All encoded features (a total of $8+4\times(N+1)$) are concatenated and passed through a stack of fully connected (FC) layers \citep{1986Natur.323..533R}. The first three layers ($FC_1$--$FC_3$) use Leaky ReLU activations to aid optimization and reduce saturation; each hidden layer contains 128 neurons.
Overall, the explicit encodings---polynomial for magnitude and color, Fourier for position---allow the network to capture subtle feature-dependent variations.


The star counts are highly imbalanced as a function of instrumental magnitude, especially at the bright end, which can degrade performance in that regime. To mitigate this, we randomly split the data into training and test sets in an 8:2 ratio and resample the training set. Specifically, we group training stars into 20 magnitude bins and up-sample (with replacement) bins below the mean count to the mean, leaving the remaining bins unchanged.
We adopt the Huber loss (Eq.~\ref{eq:6}) with transition threshold $\delta=0.2$, which matches the typical precision of MYX25. The Huber loss combines the advantages of mean squared error (MSE) and mean absolute error (MAE): it remains sensitive to small errors while being robust to outliers.

\begin{align}
\mathcal{L}_{\text{Huber}}(y, \hat{y}) = 
\begin{cases}
\frac{1}{2}(y - \hat{y})^2, & \text{if } |y - \hat{y}| \leq 0.2 \\
0.2 \cdot |y - \hat{y}| - 0.02, & \text{otherwise}
\end{cases}
\label{eq:6} 
\end{align}

We train using the Adam optimizer (initial learning rate $2\times10^{-4}$; batch size 1{,}000) with an adaptive decay schedule. Every 20 epochs, we evaluate on the test set; if the mean residual over all sources falls below 0.025 and the test loss has decreased by less than $2\times10^{-4}$ since the previous evaluation, we reduce the learning rate by a factor of 0.2. Training terminates once the learning rate falls below $2\times10^{-6}$. This schedule mitigates overfitting and yields approximately normally distributed calibration residuals.

All hyperparameters were selected empirically on our dataset. We find that MFF-Net is stable over a reasonable range of settings. Moreover, repeated independent calibrations of the same plate achieve consistent precision (fluctuations $<0.01$ mag), indicating robustness to random initialization.

\begin{figure*}[htbp!] \centering
\includegraphics[width=17cm]{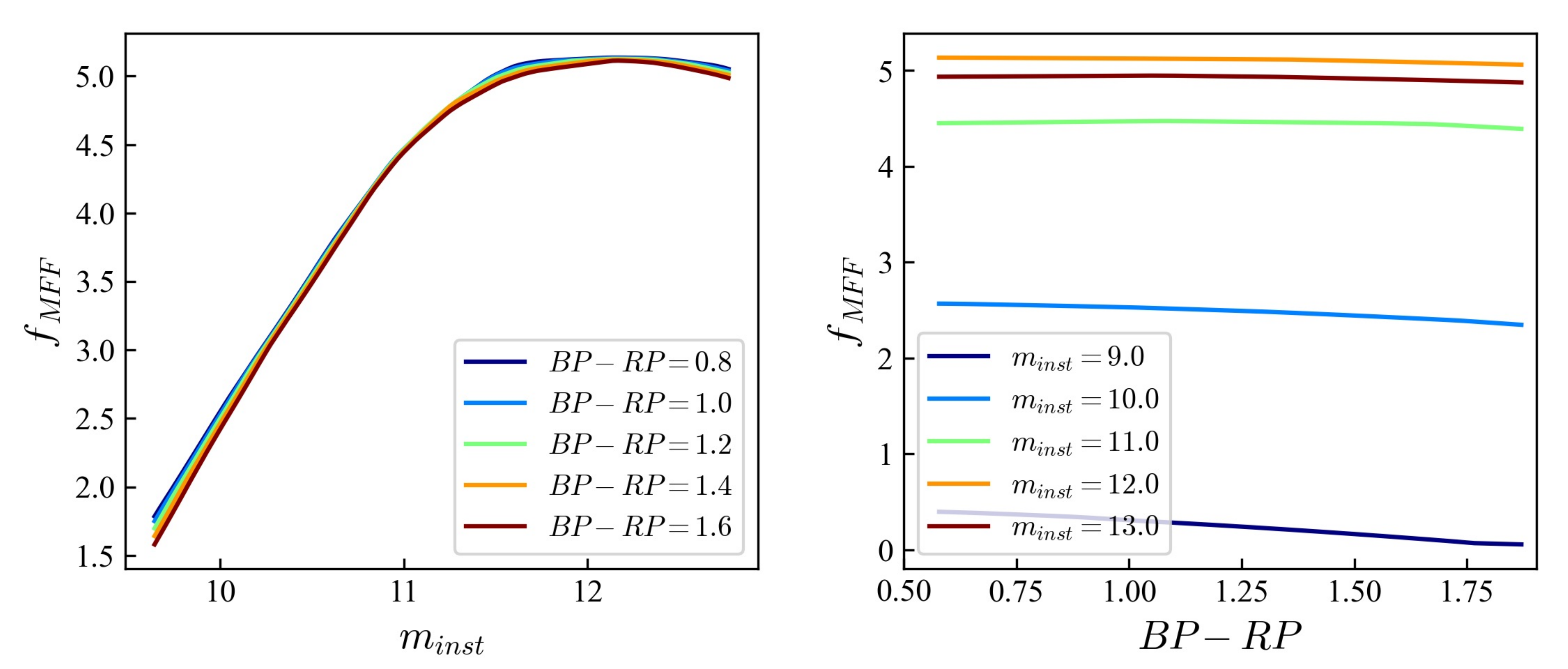} 
\caption{MFF-Net correction as a function of magnitude (left) and color (right), evaluated at the field center of plate ZT5810N973001. In the left panel, curves correspond to different colors; in the right panel, curves correspond to different magnitudes.}
\label{Fig:macitem}
\end{figure*}

\begin{figure*}[htbp!] \centering
\includegraphics[width=18cm]{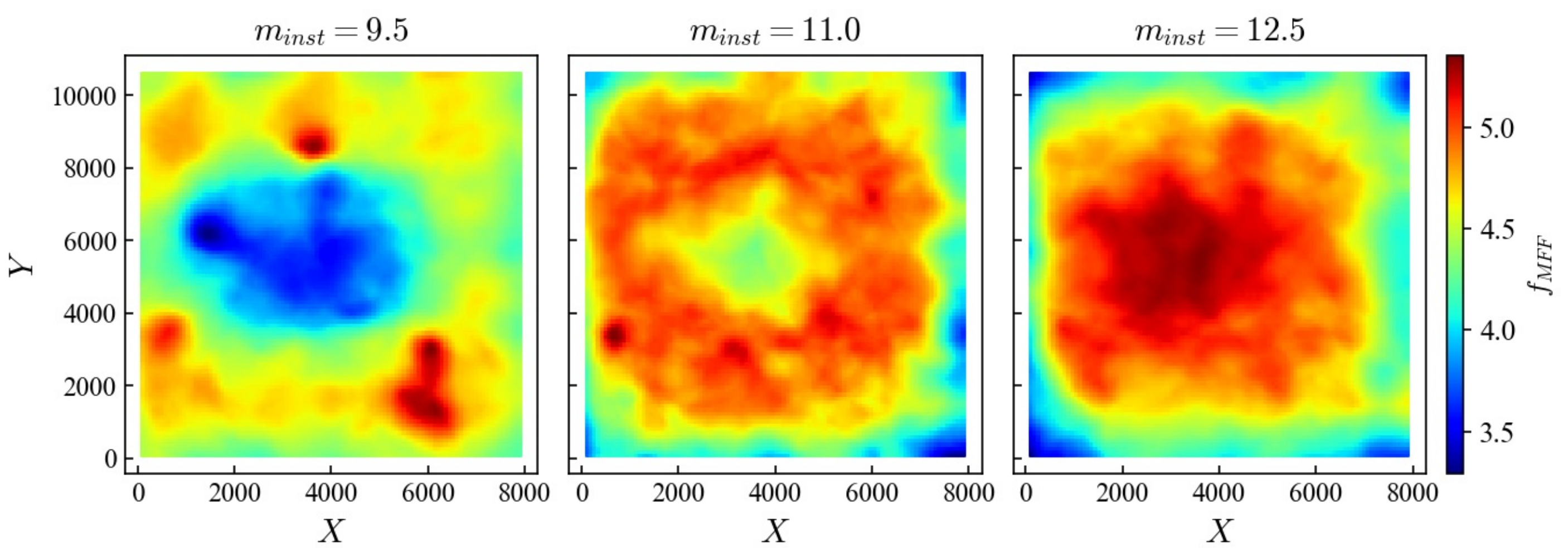}
\caption{MFF-Net correction as a function of pixel coordinates for plate ZT5810N973001, shown for stars of different magnitudes at fixed $BP-RP=1$.}
\label{Fig:flatitem}
\end{figure*}

\begin{figure}[htbp!] \centering
\includegraphics[width=8.5cm]{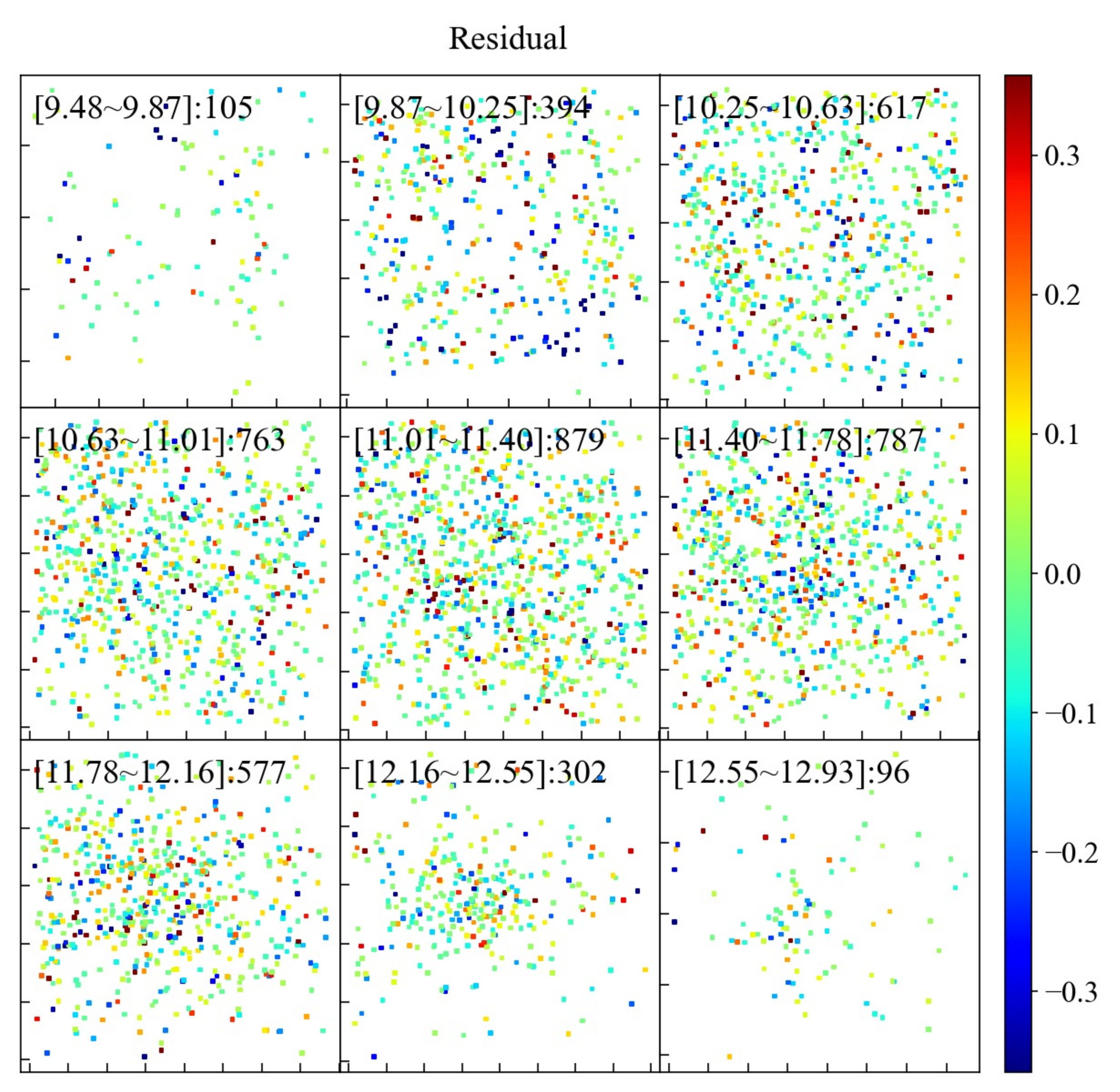} 
\caption{MFF-Net calibration residual of ZT5810N973001, plotted as in the left panel of Fig.~\ref{Fig:couple_mrmt}.}
\label{Fig:couple_net}
\end{figure}

We apply MFF-Net to plate ZT5810N973001 with $N=2$. Figure~\ref{Fig:macitem} shows the learned correction at the plate center as a function of magnitude and color, while Figure~\ref{Fig:flatitem} shows the corresponding spatial corrections at different magnitudes.
The magnitude--color coupling is present but substantially weaker than the magnitude--position coupling. 
Fig.~\ref{Fig:residual} (lower panels) shows the post-calibration residuals versus instrumental magnitude, color, and position, and Fig.~\ref{Fig:couple_net} shows their spatial distribution in nine magnitude bins. The residuals are approximately uniform both overall and within each bin. 
Compared to the MYX25 results (Fig.~\ref{Fig:couple_mrmt}, left), MFF-Net effectively removes the brightness--position coupling and improves the photometric precision by nearly a factor of two, from 0.24~mag to 0.12~mag. This demonstrates that the network can capture nonlinear, coupled dependencies among magnitude, color, and position, thereby suppressing spurious structure and enabling a more complete correction.

\begin{figure*}[htbp!] \centering
\includegraphics[width=18cm]{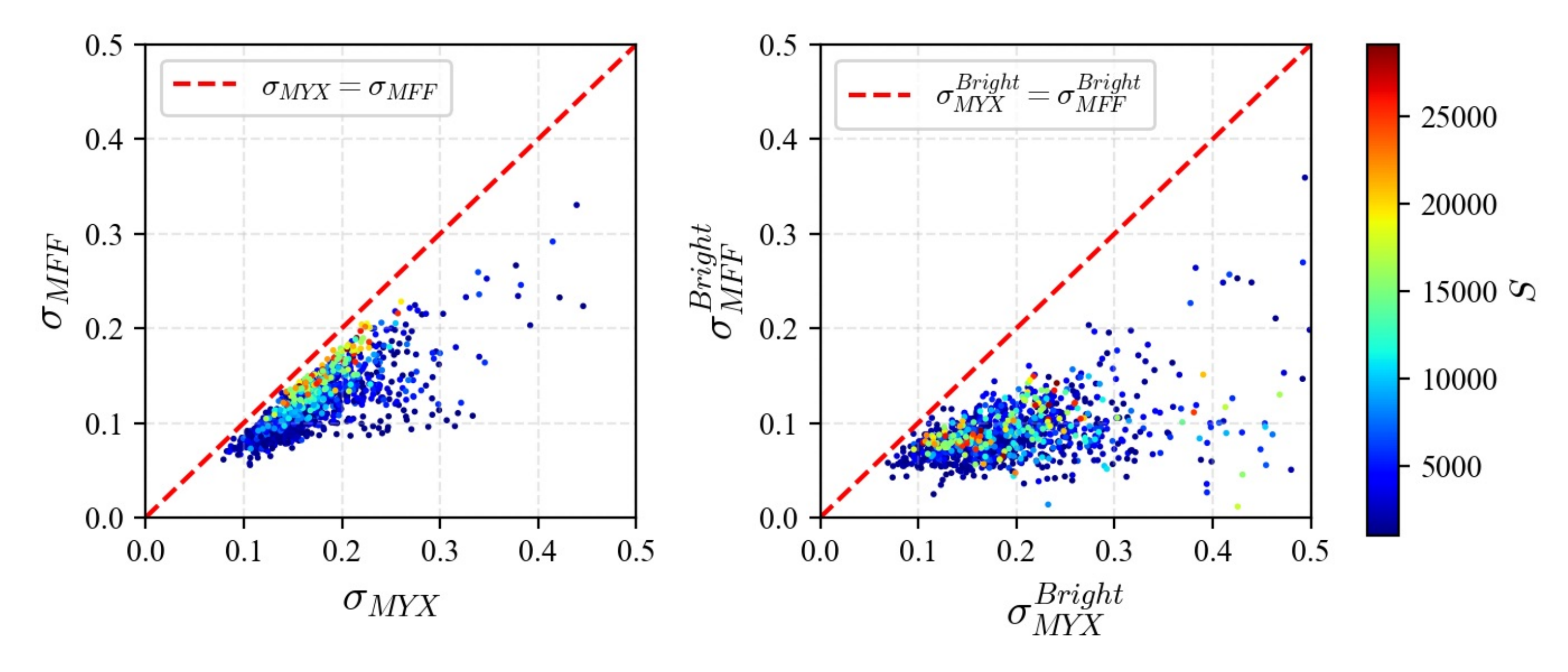}
\caption{Photometric precision comparison (in magnitudes) for 1,200 plates (each with $>1,000$ reference stars). $\sigma_{\mathrm{MYX}}$: MYX25 on all sources; $\sigma_{\mathrm{MFF}}$: MFF-Net on all sources; $\sigma^{\mathrm{Bright}}_{\mathrm{MFF}}$: MFF-Net on bright sources only; $\sigma^{\mathrm{Bright}}_{\mathrm{MYX}}$: MYX25 on bright sources only. Bright sources satisfy $m_{\mathrm{inst}} < (m_{\min}+m_{\max})/2$, where $m_{\min}$ and $m_{\max}$ are the minimum and maximum instrumental magnitudes on each plate. Colors encode the number of reference stars per plate. The dashed line denotes the 1:1 relation; points below it indicate higher precision with MFF-Net.}
\label{Fig:batch}
\end{figure*}

To further assess MFF-Net, we select 1{,}200 Chinese plates with $>1{,}000$ standard stars, using stratified sampling in reference-star count. As shown in Figure~\ref{Fig:batch} (left), MFF-Net yields higher photometric precision than MYX25 for all plates. Over the 5th--95th percentile range, the precision improves from 0.11--0.26~mag (MYX25) to 0.08--0.18~mag (MFF-Net).

Across this sample, image-quality patterns vary widely (e.g., annular versus linear features, and strong versus weak small-scale structure), and stellar spatial distributions range from nearly uniform to highly inhomogeneous. MFF-Net remains robust across these conditions, with the largest gains on plates where image quality varies most strongly with magnitude.

Figure~\ref{Fig:batch} (left) also indicates that MFF-Net's improvement over MYX25 decreases as the number of reference stars increases. The reason is that plates with many reference stars are dominated by faint sources: MYX25's flat-field term is therefore constrained mainly by the faint end, and bright-end errors contribute little to the overall precision.

When we evaluate MFF-Net on bright sources only (Figure~\ref{Fig:batch}, right), a different pattern emerges: plates with many reference stars no longer cluster near the 1:1 line, but instead lie significantly below it. This indicates that MFF-Net's bright-end advantage persists even when the total reference-star count is high.
MFF-Net reaches a bright-end precision of $\sim 0.1$~mag, about a factor of two better than MYX25 on bright sources.
Therefore, although the overall gain on star-rich plates can appear modest because bright sources are relatively sparse, the improvement in bright-end calibration remains particularly valuable.

For these 1{,}200 test plates, training on a single RTX~4090 GPU takes from a few tens of seconds to just over 100~s per plate, increasing with the number of reference stars.


\section{Conclusion and Outlook} \label{sec:conclusion}

In this study, analysis of the MYX25-calibrated catalogs for Chinese digitized plates shows that the residuals retain a magnitude-dependent spatial structure. Through numerical simulations, we attribute this behavior to magnitude-dependent spatial variations in image quality, a common characteristic of photographic plates.

To address these limitations, we developed the MFF-Net that jointly models the coupled, nonlinear dependencies among magnitude, color, and position, thereby reducing parameter coupling and improving photometric calibration accuracy. Tests on 1{,}200 digitized Chinese plates show that MFF-Net consistently outperforms MYX25, improving the 5th--95th percentile precision from 0.11--0.26~mag to 0.08--0.18~mag. Although the overall gain decreases on plates with many reference stars, the improvement for bright sources remains substantial (about a factor of two over MYX25), which is critical for bright-end photometric precision.

MFF-Net provides a practical framework for recalibrating historical photographic plate archives and improving their scientific utility.
In future work, we will apply MFF-Net to update the Chinese digitized-plate catalogs and adapt the approach to other plate archives.



\newpage 
This work is supported by the National Natural Science Foundation of China (NSFC 12222301, 12173007) and the National Key Basic R\&D Program of China (2024YFA1611901, 2024YFA1611601). L.Y. is supported by the Beijing Natural Science Foundation (No. 1264072); K.X. by the NSFC (No. 12403024); B.H. by the NSFC (No. 124B2055); and Y.Y. by the NSFC (No. 12473070).


\end{document}